\documentclass[conference,a4paper]{IEEEtran}
\usepackage[utf8]{inputenc}
\usepackage{lipsum}
\usepackage{amsmath,amssymb,amsthm}
\usepackage{color}
\usepackage[linesnumbered,ruled,vlined,titlenumbered]{algorithm2e}
\usepackage[innermargin=0.545in,outermargin=0.65in,top=0.545in,bottom=.7in]{geometry}
\usepackage{algorithmicx}
\usepackage{tikz}
\usepackage{pgfplots}
\usepackage{tikzscale}
\pgfplotsset{compat=newest}
\usepackage{color}
\usepackage{mathtools}
\usepackage{url}
\usepackage{booktabs}
\usepackage{flushend}

\newtheorem{theorem}{Theorem}
\newtheorem{definition}{Definition}
\newtheorem{lemma}{Lemma}
\newtheorem{corollary}{Corollary}

\newtheorem{remark}{Remark}

\newtheorem{example}{Example}

\def\ve#1{{\mathchoice{\mbox{\boldmath$\displaystyle #1$}}%
		{\mbox{\boldmath$\textstyle #1$}}%
		{\mbox{\boldmath$\scriptstyle #1$}}%
		{\mbox{\boldmath$\scriptscriptstyle #1$}}}}

\IEEEoverridecommandlockouts

\usepackage{color}

\author{\IEEEauthorblockN{Carmen Sippel$^1$, Cornelia Ott$^{1}$, Sven Puchinger$^{2}$, and Martin Bossert$^{1}$}
\IEEEauthorblockA{%
$^{1}$Institute of Communications Engineering, Ulm University, Germany \\
$^{2}$Institute for Communications Engineering, Technical University of Munich (TUM), Germany \\
{\tt\small \{carmen.sippel, cornelia.ott, martin.bossert\}@uni-ulm.de}, {\tt\small sven.puchinger@tum.de}
}
\thanks{This work was supported by the German Research Council (DFG) under Grants Bo 867/32 and  Bo 867/35.
This is an extended version of a paper~\cite{sippel2019RSchar0} accepted at ISIT 2019.
}%
}
\title{\LARGE \bf
Reed--Solomon Codes over Fields of Characteristic Zero
}

\newcommand{\G}{\ve{G}}
\renewcommand{\H}{\ve{H}}
\newcommand{\NN}{\mathbb{N}}
\newcommand{\ZZ}{\mathbb{Z}}
\newcommand{\QQ}{\mathbb{Q}}
\newcommand{\Code}{\mathcal{C}}

\newcommand{\CGRS}{\Code_\mathrm{GRS}}
\newcommand{\HGRS}{\H_\mathrm{GRS}}
\newcommand{\GGRS}{\G_\mathrm{GRS}}
\newcommand{\info}{\ensuremath{\ve{u}}}
\renewcommand{\c}{\ve{c}}
\renewcommand{\r}{\ve{r}}
\newcommand{\e}{\ve{e}}
\newcommand{\s}{\ve{s}}
\renewcommand{\a}{\ve{a}}
\renewcommand{\b}{\ve{b}}
\newcommand{\A}{\ve{A}}
\newcommand{\B}{\ve{B}}
\newcommand{\wtH}{\mathrm{wt}_\mathrm{H}}
\newcommand{\den}{\mathrm{den}}
\newcommand{\num}{\mathrm{num}}
\newcommand{\lcm}{\mathrm{lcm}}
\newcommand{\alphavec}{\ensuremath{\ve{\alpha}}}
\makeatletter
\newcommand{\removelatexerror}{\let\@latex@error\@gobble}
\makeatother
\newcommand{\printalgoIEEE}[1]
{{\centering
\scalebox{0.97}{
\removelatexerror
\begin{tabular}{p{\columnwidth}}
\begin{algorithm}[H]
 \begin{small}
 #1
 \end{small}
\end{algorithm}
\end{tabular}
}
}
}
\newcommand{\bigO}{\ensuremath{O}}
\newcommand{\Eset}{\mathcal{E}}
\newcommand{\smallsum}{\textstyle\sum}

\begin{document}

\maketitle
\thispagestyle{empty}
\pagestyle{empty}

\begin{abstract}
We study Reed--Solomon codes over arbitrary fields, inspired by several recent papers dealing with Gabidulin codes over fields of characteristic zero.
Over the field of rational numbers, we derive bounds on the coefficient growth during encoding and the bit complexity of decoding, which is polynomial in the code length and in the bit width of error and codeword values.
The results can be generalized to arbitrary number fields.
\end{abstract}

\section{Introduction}

Reed--Solomon (RS) codes were introduced in 1960 \cite{reed1960polynomial} and have become some of the most used classes of algebraic codes.
The codes are defined over finite fields and can be efficiently decoded, both up to half-the-minimum distance \cite{sugiyama1975method,RothBook}
and beyond \cite{sudan1997decoding,guruswami1998improved,wu2008new,schmidt2010syndrome,rosenkilde2018power}. 
RS codes over the field of complex numbers, also called \emph{complex RS codes} or \emph{analog codes}, were introduced in \cite{wolf_redundancy_1983} and \cite{marshall_coding_1984}, and their decoding was studied in \cite{henkel1989decodierung}.
Further decoding principles known from finite fields, as well as applications of the codes to compressed sensing, were analyzed and numerically evaluated in \cite{parvaresh_explicit_2008,MRZB_2015,ZoerleinDiss,mohamed2017guruswami}.

In this paper, we study RS codes over arbitrary fields and their properties.
The idea is inspired by several recent publications that have dealt with Gabidulin codes, the rank-metric analog of RS codes, over field of characteristic zero (in particular number fields), which were first described in \cite[Section~6]{roth1996tensor}, \cite{augot2013rank,robert2016quadratic,muelich2016alternative,augot2017generalized}.
These codes have applications in space-time coding \cite{robert2015new} and low-rank matrix recovery \cite{muelich2017low}.
Among other possible applications, we believe that RS codes over number fields (in particular over $\QQ$ and $\QQ[\mathrm{i}]$)
are suitable to replace complex RS codes in some applications, e.g., the compressed sensing scenario studied in \cite{MRZB_2015,ZoerleinDiss}, since there are no numerical issues.
Studying these applications goes beyond the scope of this paper and needs to be done in future work.

For decoding, we adapt a syndrome-based half-the-minimum-distance decoder known from finite fields and restrict to the field of rational numbers, which is an infinite field and an exact computation domain. 
In contrast to complex RS codes, we face the problem of coefficient growth (i.e., large numerators and denominators, also known as intermediate expression swell) during computations instead of numerical issues caused by floating point operations. This substantially influences the bit complexity of decoding algorithms compared to finite fields, where field elements can be represented with a fixed number of bits and field operations cost a constant number of bit operations (for a given field size).
On the other hand, there are no numerical problems, in contrast to complex RS codes.

We derive bounds on the coefficient growth during encoding and decoding. This implies an upper bound on the bit complexity of decoding, which is polynomial in the code length and bit width of the error values.

The results can be extended to more classes of number fields, for instance $\QQ[\mathrm{i}]$, or cyclotomic and Kummer extensions.
The adaption is technical, which is why we restrict ourselves to $\QQ$ here.
Furthermore, it is reasonable to expect that our results can be used to analyze the runtimes of the existing decoders for Gabidulin codes over number fields \cite{augot2013rank,robert2016quadratic,muelich2016alternative} in bit operations, instead of field operations (which do not consider coefficient growth). 

\section{Reed--Solomon Codes over Arbitrary Fields}

The following is a straight-forward generalization of generalized Reed--Solomon (GRS) codes to arbitrary, possibly infinite, fields.
We define the codes as in \cite{RothBook} by simply replacing the finite field by an arbitrary field $K$.

\begin{definition}\label{def:GRS_codes}
Let $k$ and $n$ be integers such that $k<n$.
Furthermore, choose $\alpha_1,\dots,\alpha_n \in K$ to be distinct non-zero elements of the field $K$, and $v_1,\dots,v_n$ to be non-zero elements from $K$. 
The corresponding \emph{generalized Reed--Solomon (GRS) code} is defined by the linear code $\CGRS \subseteq K^n$ with parity check matrix
\begin{equation*}
\setlength{\arraycolsep}{1pt}
\HGRS = \begin{pmatrix}
1 & 1 & \dots & 1 \\
\alpha_1 & \alpha_2 & \dots & \alpha_n \\
\vdots & \vdots & \dots & \vdots \\
\alpha_1^{n-k-1} & \alpha_2^{n-k-1} & \dots & \alpha_n^{n-k-1}
\end{pmatrix}
\begin{pmatrix}
v_1 & & & \\
    & v_2 & & \\
    &     & \ddots & \\
    &     &        & v_n
\end{pmatrix} .
\end{equation*}
\end{definition}

The proof that RS codes are MDS is straightforward using the same arguments as in the finite field case (see, e.g., \cite{RothBook}).

\begin{theorem}
Any GRS code is MDS, i.e., $d = n-k+1$.
\end{theorem}

\begin{theorem}\label{thm:G_GRS}
The code $\CGRS$ as defined in Definition~\ref{def:GRS_codes} has a generator matrix of the form
\begin{equation*}
\setlength{\arraycolsep}{4pt}
\GGRS = \begin{pmatrix}
1 & 1 & \dots & 1 \\
\alpha_1 & \alpha_2 & \dots & \alpha_n \\
\vdots & \vdots & \dots & \vdots \\
\alpha_1^{k-1} & \alpha_2^{k-1} & \dots & \alpha_n^{k-1}
\end{pmatrix}
\begin{pmatrix}
v_1' & & & \\
    & v_2' & & \\
    &     & \ddots & \\
    &     &        & v_n'
\end{pmatrix}\, ,
\end{equation*}
where the $\alpha_i$ are the same as in Definition~\ref{def:GRS_codes} and the $v_i'$ are non-zero elements of $K$, given by the following linear system of equations:
\begin{equation}
\sum_{i=1}^{n} \alpha_i^r v_i v_i' = 0 \quad \forall \, r = 0,\dots,n-2 \, . \label{eq:v_iv_i'_system}
\end{equation}
\end{theorem}

\begin{IEEEproof}
The proof is analog to \cite[Proposition~5.2]{RothBook}.
\end{IEEEproof}

\begin{theorem}\label{thm:closed_form_v_iv_i'}
Fix $\alpha_1,\dots,\alpha_n \in K \setminus \{0\}$ distinct. Let $v_i,v_i'$ be non-zero such that
\begin{equation*}
v_i v_i' = \textstyle\prod_{\substack{1\leq j \leq n\\i\neq j}} (\alpha_i - \alpha_j)^{-1} =: w_i \ . 
\end{equation*}
Then, the corresponding matrices $\GGRS$ and $\HGRS$ are generator and parity check matrix of the same code, respectively.
\end{theorem}

\begin{IEEEproof}
If we solve the system in \eqref{eq:v_iv_i'_system} for $v_i v_i'$, then the solution is given by a non-zero element in the kernel of a Vandermonde matrix $V_{n,n-1}$ with $n-1$ rows. Such a vector is given by $w_i$ as above, since $(w_1,\dots,w_n)^\top$ is the last column of the inverse Vandermonde matrix $V_{n,n}$ (cf.~\cite[eq.~(9),(10): Set $k=\mu=n$, note that $\sigma_{0,n-1}^i=1$.]{macon1958inverses}). 
\end{IEEEproof}

\begin{remark}
Theorem~\ref{thm:closed_form_v_iv_i'} implies a method to compute the $v_i$ from the $v_i'$ or vice-versa, i.e.,
\begin{equation*}
v_i = \frac{w_i}{v_i'} \quad \text{and} \quad v_i' = \frac{w_i}{v_i}
\end{equation*}
\end{remark}

\section{Coefficient Growth over the Rational Numbers}

In the following, we study the coefficient growth during computations over the rational numbers. The results can in principle be extended to a wider class of number fields by generalizing the following well-known notion of bit width.

\begin{definition}[Generalization of {\cite[p.~142]{von2013modern}}]\label{def:bit width}
Let $a$ be an element of one of the sets in $\{\mathbb{Z}, \mathbb{Q}, \mathbb{Q}[x], \mathbb{Q}^{k\times n}\}$. We define its \emph{bit width} $\lambda(a)$ as follows:
\begin{itemize}
\item $a \in\mathbb{Z}$:
\begin{equation*}
\lambda(a) \coloneqq
\begin{cases}
\lfloor \log_2(|a|)\rfloor +1, &\text{if } a \neq 0 \\
0, &\text{if } a = 0 
\end{cases}
\end{equation*}
\item $a= \frac{b}{c} \in \mathbb{Q}$ with $b,c \in \mathbb{Z}$, $c \neq 0$, and $\gcd(b,c)=1$:
\begin{equation*}
\lambda(a) \coloneqq \max \{\lambda(b), \lambda(c)\}.
\end{equation*}
\item $a(x)=\sum_{i=0}^n \frac{a_i}{b}\cdot x^i \in \mathbb{Q}[x]$ with $a_i\in \mathbb{Z}$ and $b\in \mathbb{N}\setminus\{0\}$ such that $\gcd(a_0, \ldots, a_n, b)=1$:
\begin{equation*}
\lambda(a(x)) \coloneqq \max\{\lambda(a_0), \ldots, \lambda(a_n), \lambda(b)\}.
\end{equation*}
\item $\A =(a_{ij})\in \mathbb{Q}^{k\times n}$:
\begin{equation*}
\lambda(\A)=\max \{ \lambda(a_{ij}) \, : \, i=1,\ldots ,k \text{ and }  j=1,\ldots,n\}.
\end{equation*}
\end{itemize}
\end{definition}

\begin{remark}
Note that the bit width does not necessarily cover the required memory space, e.g.\ for a rational number only the maximum of the bit widths of numerator and denominator is regarded, which means, that for equal bit widths of numerator and denominator the actual memory space is twice the bit width.
\end{remark}

\begin{theorem}[Coefficient Growth {\cite[p.~142]{von2013modern}}] \label{thm:recap}
Let $a,b$ be polynomials in $\mathbb{Z}[x]$ with coefficients $a_i$, $b_j$ for $i \in \{0, \ldots, n\coloneqq \deg(a(x))\}$ and  $j \in \{0, \ldots, m\coloneqq\deg(b(x))\}$ and $c,d \in \mathbb{Q}$, then the following statements hold:
\begin{enumerate}
\item $\lambda(a(x)+b(x))\leq \max\{\lambda(a(x)), \lambda(b(x))\}+1$
\item\label{itm:lambda_poly_mult} $\lambda(a(x)\cdot b(x)) \leq \lambda(a(x)) + \lambda(b(x)) + \lambda(\min\{n,m\} +1)$
\item $\lambda(cd)\leq\lambda(c)+\lambda(d)$
\item $\lambda(\frac{c}{d})\leq\lambda(c)+\lambda(d)$
\item $\lambda(c+d)\leq \lambda(c)+\lambda(d)+1$
\end{enumerate}

\begin{IEEEproof}
1) For polynomials $a,b$ of degree $0$, it's trivial.
For polynomials $a,b$ of arbitrary degree (w.l.o.g. we assume $n\geq m$) with coefficients $a_i$, $b_j$ for $i \in \{0, \ldots, n\}$ and  $j \in \{0, \ldots, m\}$ it holds that
\begin{align*}
&\lambda(a(x)+b(x))\\
=& \max\{\lambda(a_0+b_0), \ldots, \lambda(a_m+b_m), \lambda(a_{m+1}), \ldots, \lambda(a_n)\}\\
\leq &\max\{\max\{\lambda(a_0), \lambda(b_0)\}+1, \ldots, \max\{\lambda(a_m),\lambda(b_m)\}\\&+1, \lambda(a_{m+1}), \ldots, \lambda(a_n)\}\}\\
\leq &\max\{\lambda(a_0), \ldots, \lambda(a_n), \lambda(b_0), \ldots, \lambda(b_n)\}+1\\
=&\max\{\max\{\lambda(a_0), \ldots, \lambda(a_n)\}, \\ &\max\{\lambda(b_0, \ldots, \lambda(b_n))\}\}+1\\
=&\max\{\lambda(a(x)), \lambda(b(x))\}+1
\end{align*}
2)\begin{align*}
&\lambda(a(x)\cdot b(x))=\max_{k\in\{0, \ldots, n+m\}}\{\lambda(\sum_{\substack{(i,j):i+j=k\\0\leq i\leq n\\ 0\leq j\leq m}}a_ib_j)\}\\
\leq &\lambda(\max\{a_0, \ldots, a_{n}\}\cdot \max\{b_0, \ldots, b_{m)}\}\cdot (\min\{n, m\}+1))\\
=& \lfloor\log_2(|\max\{a_0, \ldots, a_{n}\}\cdot \max\{b_0, \ldots, b_{m)}\}\\&\cdot (\min\{n, m\}+1)|)\rfloor 
+1\\
\leq &\lfloor\log_2(|\max\{a_0, \ldots, a_{n}\}|)\rfloor + 1 \\&+ \lfloor\log_2(|\max\{b_0, \ldots, b_{m)}\}|)\rfloor +1 \\&+ \lfloor\log_2(|(\min\{n, m\}+1)|)\rfloor+1\\
=&\lambda(\max\{a_0, \ldots, a_{n}\})+ \lambda(\max\{b_0, \ldots, b_{m)}\})\\&+\lambda(\min\{n, m\}+1)\\
=&\lambda(a(x))+\lambda(b(x))+\lambda(\min\{n, m\}+1)
 \end{align*}
 3)We write $c=\frac{c_1}{c_2}$ and $d=\frac{d_1}{d_2}$ with $c_1, c_2,d_1,d_2 \in \mathbb{Z}$ with $\gcd(c_1,c_2)=\gcd(d_1,d_2)=1$. It holds:
 \begin{align*}
  &\lambda(cd)=\lambda\left(\frac{c_1\cdot d_1}{c_2\cdot d_2}\right)\leq \max\{\lambda(c_1 \cdot d_1), \lambda(c_2 \cdot d_2)\}
  \\
 =&\max\{\lfloor \log_2(|c_1 \cdot d_1|)\rfloor+1, \lfloor\log_2(|c_2 \cdot d_2|)\rfloor+1\}\\
\leq  &\max\{\lfloor \log_2(|c_1|)\rfloor+1 + \lfloor \log_2(|d_1|)\rfloor+1, \\
&\lfloor\log_2(|c_2|)\rfloor+1+ \lfloor\log_2(|d_2|)\rfloor+1\}\\
  \leq  &\max \{\lambda(c_1)+\lambda(d_1), \lambda(c_2)+\lambda(d_2)\}\\
  \leq&\lambda(c)+\lambda(d).
 \end{align*}
4) 
Let $c' = c^{-1}$, then the claim follows by 3).\\
 5) \begin{align*}
& \lambda(c+d)=\lambda\left(\frac{c_1 d_2 + d_1 c_2}{c_2 d_2}\right)\\
=&\max\{\lambda(c_1 d_2 + d_1 c_2),\lambda(c_2 d_2)\}\\
\leq & \max\{ \max\{\lambda(c_1d_2), \lambda(c_2d_1)\}+1,\lambda(c_2 d_2)\}\\
=&\max\{\lambda(c_1d_1)+1,\lambda(d_2c_2)+1, \lambda(c_2d_2)\}\\
\leq&\lambda(c)+\lambda(d)+1 \hspace{6cm} \IEEEQEDhere
 \end{align*}
\end{IEEEproof}

\end{theorem}
Theorem~\ref{thm:recap} implies the following statements about coefficient growth in a vector or matrix multiplication.
\begin{theorem}[Multiplication of vector and vector]\label{thm:lambda_vector-vector_product}
Let $\a,\b \in \QQ^n$. Then,
\begin{equation*}
\lambda(\a \b^\top) \leq n \cdot (\lambda(\a)+\lambda(\b) + 1).
\end{equation*}
\end{theorem}

\begin{IEEEproof}
Using Theorem~\ref{thm:recap}, we obtain
\begin{align*}
\lambda(\a \b^\top) &= \lambda\left( \smallsum_{i=1}^{n} a_i b_i \right) \leq \smallsum_{i=1}^n\left( \lambda(a_ib_i) \right) + n - 1 \\
					&\leq \smallsum_{i=1}^n\left( \lambda(a_i) + \lambda(b_i) \right) + n - 1, 
\end{align*}
which implies the claim.
\end{IEEEproof}

\begin{theorem}[Multiplication of vector and matrix]\label{thm:lambda_vector-matrix_product}
Let $\A \in \QQ^{n \times r}$ and $\B \in \QQ^{r \times m}$. Then,
\begin{equation*}
\lambda(\A \B) \leq r(\lambda(\A)+\lambda(\B) + 1).
\end{equation*}
\end{theorem}

\begin{IEEEproof}
The statement directly follows from Theorem~\ref{thm:lambda_vector-vector_product} since any entry of the product $\A \B$ is the result of the multiplication of a row of $\A$ with a column of $\B$, which are both vectors of length $r$.
\end{IEEEproof}

\section{Coefficient Growth in Encoding RS Codes over the Rational Numbers}

In this section we study the coefficient growth during encoding with a generator matrix $\G$, i.e., we derive a bound on the bit width of the codeword $\c \in \QQ^n$ obtained from an information word $\info \in \QQ^k$ with a given bit width.
We also show how to reduce the coefficient growth compared to the standard generator matrix given in Theorem~\ref{thm:G_GRS}.

\begin{theorem} \label{thm:lambda_c}
	Let $\c$ be an RS codeword generated by encoding $\info \in \QQ^{k}$ with generator matrix $\G \in \QQ^{k\times n}$. Then
	\begin{align*}
		\lambda(\c) \leq k(\lambda(\info) + \lambda(\G) + 1).
	\end{align*}
\end{theorem}

\begin{IEEEproof}
	The claim follows directly from Theorem~\ref{thm:lambda_vector-matrix_product}.
\end{IEEEproof}

Hence, the maximal coefficient growth depends heavily on the choice of the generator matrix. We can bound $\lambda(\G)$ as follows.

\begin{corollary} \label{cor:GVandermonde}
	Let $\G$ be a generator matrix as in Theorem~\ref{thm:G_GRS} using $\alphavec = (\alpha_1,\dots,\alpha_n)$ and $\ve{v}' = (v_1',\dots,v_n')$. Then  
	\begin{align} \label{eq:ineqG}
		\lambda(\G) \leq (k-1)\lambda(\alphavec) + \lambda(\ve{v}').
	\end{align}	
\end{corollary}

\begin{IEEEproof}
We have
	\begin{align*}
		\lambda(\G) &\underset{\text{Def. }\ref{def:bit width}}{=} \max\limits_{\substack{i=1,\dots,n\\j=0,\dots,k-1}}\{ \lambda(\alpha_i^j v_i') : \alpha_i,v_i' \in \QQ\} \\
		& \underset{\text{Th. }\ref{thm:recap}}{\leq} \max\limits_{\substack{i=1,\dots,n\\j=0,\dots,k-1}} \{ j \lambda(\alpha_i) + \lambda(v_i') : \alpha_i,v_i' \in \QQ \}, 
	\end{align*}
which implies the claim.
\end{IEEEproof}

\begin{remark}
	For $v_i' = 1,\, i=1,\dots,n$ the addition of $\lambda(\ve{v}')=1$ can be omitted, since then $\alpha_i^j v_i' = \alpha_i^j \, \forall i$.
	In this case, \eqref{eq:ineqG} is fulfilled with equality.
\end{remark}

\begin{example}
	If we choose $\alphavec = (1,\dots,n)$, $\ve{v}'=(1,\dots,1)$ then $\lambda(\G) = (k-1)\lambda(n)$, whereas the allocated memory is
	\begin{align*}
		&\sum_{j=1}^{k-1} j \sum_{i=1}^n (\lfloor \log_2(i) \rfloor + 1) \geq \, \sum_{j=1}^{k-1} j \sum_{i=1}^n \log_2(i) \\
&=  \frac{(k-1)k}{2} \log_2(n!) = (k-1) \frac{k}{2} \lambda(n!) \\
&\geq (k-1) \lambda(n) \quad \text{for } k>2.
	\end{align*}
	This example shows, that the bit width does not cover the necessary memory space.
\end{example}

According to \cite[Theorem~1]{Roth:1985:GMM:7030.7044} the generator matrix of a GRS code can be brought to systematic form, where the identity matrix is followed by a Cauchy matrix, which is defined as follows. 

\begin{table*}
	\caption{Upper bounds for the bit widths of generator and parity check matrix for several choices for $v_i$ and $v_i'$.} \label{tab:choicesv}
	\vspace{-0.5cm}
	\begin{center}
	\begin{tabular}{p{3cm}p{4cm}p{4cm}p{4cm}}
		\toprule
							& $\lambda(\GGRS)$ 	& $\lambda(\G_\mathrm{Cauchy})$ & $\lambda(\HGRS)$  \\
		\midrule
		general 			& $(k-1)\lambda(\alphavec) + \lambda(\ve{v}')$ 
												& $2(2k-1)\lambda(\alphavec) + 2\lambda(\ve{v}') + 2k-1$
																				& $(n-k-1)\lambda(\alphavec) +\lambda(\ve{v}) $ \\
		\midrule
		$v_i'=1$, $v_i=w_i$	&$(k-1)\lambda(\alphavec)$
												& $ 2(k-1)(2\lambda(\alphavec) + 1)$ 
																				&$(3n-k-3)\lambda(\alphavec)+n-1$\\
		$c_id_j=1$			& $(k-1)(3\lambda(\alphavec) + 1)$
												& $2\lambda(\alphavec)+1$ 		& $(3(n-k)-1)\lambda(\alphavec)+n-k$\\

		$v_i=1$, $v_i'=w_i$	& $(2n+k-3)\lambda(\alphavec)+ n-1$ 
												& $(2n-2k+1)(2\lambda(\alphavec) + 1)$ 
																				& $ (n-k-1)\lambda(\alphavec)$ \\
		\bottomrule
	\end{tabular}
	\end{center}
\vspace{-0.5cm}
\end{table*}

\begin{theorem}[Cauchy Generator Matrices~{\cite[Theorem~1]{Roth:1985:GMM:7030.7044}}] \label{thm:RothGRSCauchy}
	Let $\CGRS$ as defined in Definition~\ref{def:GRS_codes}. The code has a systematic generator matrix of the form $\G = (\ve{I}_{k\times k} \mid \A ) $, where $\ve{I}_{k\times k}$ is an identity matrix of size $k\times k$ and $\A = \left(\frac{c_i d_j}{a_i - b_j}\right)$ is a Cauchy matrix with
	\begin{align}
	a_i &= \alpha_i, & & i=1,\dots,k \label{eq:a_i}\\
	b_j &= \alpha_{j+k}, & & j=1,\dots,n-k \\
	c_i &= (v_i')^{-1} \prod_{\substack{1\leq t\leq k\\t \neq i}} (\alpha_i - \alpha_t)^{-1}, & &i=1,\dots,k \label{eq:c_i}\\
	d_j &= v_{j+k}' \prod_{1\leq t \leq k} (\alpha_{j+k} - \alpha_t), & & j=1,\dots,n-k.\label{eq:d_j}
	\end{align}
\end{theorem}
Using a generator matrix in systematic form, we obtain a generator matrix with a lower $\lambda(\G)$ than a generator matrix in Vandermonde form.

\begin{corollary}\label{cor:c_id_j=1}
Let $\G = (\ve{I}_{k\times k} \mid \A ) $ be a generator matrix of $\CGRS$ as defined in Theorem~\ref{thm:RothGRSCauchy}, where the $v_i'$ are chosen such that $c_id_j=1$ are zero, we get
\begin{align*}
\lambda(\G)\leq 2\lambda(\alphavec) +1.
\end{align*}
\end{corollary}

\begin{IEEEproof}
The claim directly follows from the fact that in this case $\A=(\tfrac{1}{\alpha_i-\alpha_{j+k}})$ and Theorem~\ref{thm:recap}.
\end{IEEEproof}

We obtain the $v_i'$ for Corollary~\ref{cor:c_id_j=1} as follows.
\begin{remark}\label{rem:c_id_j=1}
	If $v_i$ in $\HGRS$ have the following form
	\begin{align*}
	v_i = 
	\prod\limits_{\substack{k+1\leq t \leq n\\t\neq i}} (\alpha_i - \alpha_t)^{-1}
	\end{align*}
	and $v_i' = \frac{w_i}{v_i}$ from Theorem~\ref{thm:closed_form_v_iv_i'}. Then the Cauchy matrix inside the corresponding generator matrix $\GGRS$ has the form
	\begin{align*}
	A_{ij} = \frac{1}{\alpha_i - \alpha_{j+k}},
	\end{align*}
	i.e.\ $c_i$, $d_j$ from equations~\eqref{eq:c_i} and~\eqref{eq:d_j} fulfill $c_i d_j = 1$.
\end{remark}

For arbitrary $v_i'$, we get a worse bound on $\lambda(\G)$ if the Vandermonde matrix is in systematic form.

\begin{corollary}
	If $\G = (\ve{I}_{k\times k} \mid \A ) $ be a generator matrix of $\CGRS$ as defined in Theorem~\ref{thm:RothGRSCauchy}, where  $\A = \left(\frac{c_i d_j}{a_i - b_j}\right) \in \mathbb{Q}^{k\times (n-k)}, i=1,\dots,k, j=1,\dots,n-k$, then
	\begin{align*}
	\lambda(\G)\leq 2(2k-1)\lambda(\alphavec) + 2\lambda(\ve{v}') + 2k-1.
	\end{align*}
\end{corollary}

\begin{IEEEproof} The claim directly follows when inserting equations \eqref{eq:a_i} - \eqref{eq:d_j} into the definition of the bit width for matrices (see Def.~\ref{def:bit width}) and the fact that $\lambda(1)=1$, whereas the argument of $\lambda$ in the statement never gets zero. Then we have
	\begin{align*}
	\lambda(\G) \leq& \max\limits_{\substack{i=1,\dots,k\\j=1,\dots,n-k}} \{ \lambda( \frac{v_{j+k}'}{v_i'} \prod\limits_{\substack{t=1,\dots,k\\t\neq i}} \frac{\alpha_{j+k}-\alpha_t}{\alpha_i - \alpha_t} \cdot\frac{1}{\alpha_{j+k}-\alpha_i} ) \} \\
	\leq& \max\limits_{\substack{i=1,\dots,k\\j=1,\dots,n-k}} \{ \lambda( v_i') + \lambda( v_{j+k}')\\
	&+\lambda(\prod\limits_{\substack{t=1,\dots,k\\t\neq i}} (\alpha_{j+k}-\alpha_t) ) + \lambda(\prod\limits_{\substack{t=1,\dots,k\\t\neq i}} (\alpha_i - \alpha_t)^{-1})\\&+\lambda((\alpha_{j+k}-\alpha_i)^{-1})\} \\
	\leq& 2\lambda(\ve{v}') + \max\limits_{\substack{i=1,\dots,k\\j=1,\dots,n-k}}\{ \sum_{\substack{t=1,\dots,k\\t\neq i}} \lambda(\alpha_{j+k} - \alpha_t) \\
& + \sum_{\substack{t=1,\dots,k\\t\neq i}} \underbrace{\lambda((\alpha_i - \alpha_t)^{-1})}_{=\lambda(\alpha_i - \alpha_t)}+\underbrace{\lambda((\alpha_{j+k}-\alpha_i)^{-1})}_{=\lambda(\alpha_{j+k}-\alpha_i)} \} \\
	\leq& 2 \lambda(\ve{v}') + \max\limits_{\substack{i=1,\dots,k\\j=1,\dots,n-k}}\{ \sum_{\substack{t=1,\dots,k\\t\neq i}} (\lambda(\alpha_{j+k}) + \lambda(\alpha_t) + 1)\\ 
	&+ \sum_{\substack{t=1,\dots,k\\t\neq i}} (\lambda(\alpha_i) + \lambda(\alpha_t) + 1)+\lambda(\alpha_{j+k})+\lambda(\alpha_i)\\&+1 \} \\
	\leq& 2\lambda(\ve{v}') + (k-1) (2\lambda(\alphavec)+1)\cdot 2+ 2\lambda(\alphavec)+1 \\
	=& 2\lambda(\ve{v}') + (4k-2)\lambda(\alphavec) + 2k-1. \hspace{2.6cm} \IEEEQEDhere
\end{align*}
\end{IEEEproof}
By the previous statements we can regard several cases of choices for $v_i$ and $v_i'$. The resulting upper bounds on the bit widths of the generator and parity-check matrices are summarized in Table~\ref{tab:choicesv}. 
More details about the formulas in the table can be found in the appendix. 

All derived bounds depend on the bit width of the code locators $\alpha_i$. The following theorem shows how to choose such distinct $\alpha_i$ with minimal $\lambda(\alphavec)$.
\begin{theorem} \label{thm:l(n)}
	If $\alpha_1, \dots, \alpha_n \in \mathbb{Q}\backslash \{0\}$ distinct, then $\ell=\max\limits_{i}\{ \den(\alpha_i), \operatorname{num}(\alpha_i) $, where $\operatorname{num}(\alpha_i)$ is the numerator of $\alpha_i$ and $\ell$ fulfills $n \leq 4 \sum_{i=1}^{\ell} \phi(i) -2$.
\end{theorem}

\begin{remark}
	Using $\alpha_i = i, \, i =1,\dots,n$ as evaluation points for an RS code as defined in Def.~\ref{def:GRS_codes} leads to $\lambda(\c) \in \bigO(\log n)$, whereas $\alpha_i, \, i =0,\dots,n-1$ as chosen by Theorem~\ref{thm:l(n)} for evaluation points will have $\lambda(\c) \in \bigO(\log \sqrt{n})$. So asymptotically this only differs in a constant $\frac{1}{2}$.
\end{remark}

\begin{remark}
	Since $\lambda(\alphavec) \in \bigO(\log n)$, all the entries in Table~\ref{tab:choicesv} asymptotically belong to $\bigO(n \log n)$, except for $c_i d_j=1$ for the Cauchy generator matrix, i.e.\ the case of Corollary~\ref{cor:c_id_j=1}, which is $\bigO(\log n)$. So this is asymptotically the best choice.
\end{remark}
In Figure~\ref{fig:statistics_alpha} the different choices of $\alphavec$ for Vandermonde and Cauchy generator matrices are compared. 
\begin{figure}
	\centering
	\includegraphics[width=0.45\textwidth]{statalpha2_numSim1000t100n1001000-4000.tikz}
	\caption{Statistics of bit width of codewords $\c$ for several choices of $\alphavec$. The choice according to Theorem~\ref{thm:l(n)} is sorted as follows $\alphavec = (1,-1,\frac{1}{2},-\frac{1}{2},\frac{1}{3},-\frac{1}{3},\frac{2}{3},\dots)$. For the statistics $1000$ information words of bit width $100$ have been randomly chosen, the bit width of codeword $\c$ was calculated according to Def.~\ref{def:bit width} and averaged over the number of information words. The rate was chosen by $k = \lfloor n/3 \rfloor$. Higher rates lead to a larger slope.} \label{fig:statistics_alpha}
\end{figure}

\section{Coefficient Growth in Decoding RS Codes over the Rational Numbers}

There are many decoding algorithms for GRS codes, both for decoding up to half-the-minimum distance \cite{gorenstein1961class,peterson1960encoding,sugiyama1975method,welch1986error} and beyond \cite{sudan1997decoding,guruswami1998improved,wu2008new,schmidt2010syndrome,rosenkilde2018power}.

In the following, we formulate the bounded-minimum-distance (BMD) decoder described in \cite[Chapter~6]{RothBook}, which is based on \cite{gorenstein1961class,peterson1960encoding,sugiyama1975method,forney1965decoding}, over $\QQ$ instead of a finite field (which is straightforward) and analyze its complexity (which is the involved part).
For obtaining a good complexity, we use the variant based on the extended Euclidean algorithm (EEA), which was first suggested by \cite{sugiyama1975method}.
For the core step of the algorithm, the EEA, we rely on an algorithm from \cite{von2013modern}, which is designed for small intermediate coefficient growth.

\subsection{Decoding Algorithm}

The decoding algorithm in \cite[Chapter~6]{RothBook}, which is described for finite fields there, returns the codeword $\c$ given only a received word $\r = \c + \e$. The algorithm finds the error positions $\Eset := \{i \, : \, e_i \neq 0\}$ by solving a key equation and then computes the error values $e_i$ by Forney's formula.
All proofs showing the correctness of the algorithm do not make use of the finiteness of the field.
Hence, the algorithm carries over directly and we only briefly recall its idea.

The key equation consists of the following polynomials.
\begin{itemize}
\item Error locator polynomial
\begin{equation*}
\Lambda(x) = \prod_{i \in \Eset} (1-\alpha_i x) \quad \text{(unknown at receiver)}.
\end{equation*}
\item Error evaluator polynomial
\begin{equation*}
\Omega(x) = \sum_{i \in \Eset} e_i v_i \prod_{j \in \Eset \setminus \{i\}} (1-\alpha_j x) \; \text{(unknown at receiver)}.
\end{equation*}
\item Syndrome polynomial
\begin{align*}
S(x) &= \sum_{i=0}^{d-2} s_i x^i := \sum_{i=0}^{d-2} \left( \sum_{j=1}^{n} r_j v_j \alpha_j^i \right) x^i \\
&= \sum_{i=0}^{d-2} \left( \sum_{j=1}^{n} e_j v_j \alpha_j^i \right) x^i \quad \text{(known at receiver)}.
\end{align*}
\end{itemize}
Note that $[s_0,\dots,s_{d-2}] = \r \HGRS^\top$ is the syndrome with respect to the parity-check matrix $\HGRS$ of Definition~\ref{def:GRS_codes}.

The BMD decoder then finds the error locator and error evaluator polynomial by solving the following key equation.

\begin{lemma}\label{lem:decoder_lemma_1}({Key Equation, generalization of \cite[Equations~(6.4)--(6.6)]{RothBook} to $\QQ$})
Let $S(x), \Lambda(x), \Omega(x)$ be defined as above. Then,
\begin{align*}
\Lambda(x) S(x) &\equiv \Omega(x) \mod x^{d-1} \ , \\
\deg \Omega(x) &< \deg \Lambda(x) = |\Eset| \ .
\end{align*}
\end{lemma}

\begin{IEEEproof}
The proof follows by exactly the same arguments as the statements in \cite[Section~6.3]{RothBook}, by replacing the finite field by $\QQ$.
\end{IEEEproof}

The following lemma shows how to obtain the solution $\Lambda(x),\Omega(x)$ of the key equation through the extended Euclidean algorithm $\mathrm{EEA}(a,b)$ over $\QQ[x]$, which computes for two input polynomials $a,b \in \QQ[x]$ with $\deg a \geq \deg b$ a sequence 
$(r_i,s_i,t_i)$ for $i=-1,0,1,2,\dots$ 
such that we have
\begin{align*}
r_i &= s_i a + t_i b \quad \forall \, i
\end{align*}
and $\deg r_i$ is strictly monotonically decreasing until $r_i = 0$. 
For some integer $t_\mathrm{stop} \leq \deg b$, we denote by $\mathrm{EEA}(a,b,t_\mathrm{stop})$ the variant of the EEA that returns only the unique triple $(r_h,s_h,t_h)$ for which we have
\begin{align*}
\deg r_h < t_\mathrm{stop} \leq \deg r_{h-1}.
\end{align*}

\begin{lemma}\label{lem:EEA_key_eq}
Let $\wtH(\e)\leq \lfloor \tfrac{d-1}{2}\rfloor$ and $S(x)$ be the corresponding syndrome polynomial.
Furthermore, let $(r_h,s_h,t_h) \in \QQ[x]^3$ be the output of the $\mathrm{EEA}(a,b,t_\mathrm{stop})$ with input
\begin{align*}
a = \xi \cdot x^{d-1}, \quad b = \xi \cdot S(x), \quad \text{and} \quad t_\mathrm{stop} = \tfrac{d-1}{2},
\end{align*}
where $\xi \in \QQ$ is the smallest positive integer, s.t.\ $b$ has integer coefficients (i.e.\, it is the lcm of the denominators of coefficients of S). Then, we have
\begin{align*}
t_h = c \cdot \Lambda(x) \quad \text{and} \quad r_h = c \cdot \Omega(x),
\end{align*}
where $c \in \QQ$ is a non-zero constant.
\end{lemma}

\begin{IEEEproof}
The proof follows from the same arguments that lead to \cite[Corollary~6.5]{RothBook} since no step assumes the finiteness of the underlying field. The only difference besides the different field is the non-zero constant $\xi$ here, which we will use for complexity reasons below, and which simply multiplies each $r_i$, $i\geq 0$, in the EEA by $\xi$. \footnote{Note that this was stated wrongly in the short version~\cite{sippel2019RSchar0}.}
\end{IEEEproof}

The constant $c$ in Lemma~\ref{lem:EEA_key_eq} can be determined as the $0^\mathrm{th}$ coefficient of $t_h$ since $\Lambda_0 = 1$ by definition.

\begin{lemma}[Forney's formula]\label{lem:forney}
Let $\Lambda'(x) = \sum_{i>0} i \Lambda_i x^{i-1}$ be the formal derivative of $\Lambda(x)$. Then, we can compute the entries $e_i$ of the error vector $\e$ by
\begin{align*}
e_i = -\tfrac{\alpha_i}{v_i} \tfrac{\Omega(\alpha_i^{-1})}{\Lambda'(\alpha_i^{-1})}
\end{align*}
for all $i=1,\dots,n$.
\end{lemma}

\begin{IEEEproof}
The proof follows by the same arguments as in \cite[Section~6.5]{RothBook}.
\end{IEEEproof}

Note that we get $e_i = 0$ for $i\notin \Eset$, so there is no need to find $\Eset$ separately. However, this can be done to reduce the runtime of the algorithm.

\subsection{Complexity Estimation}

For bounding the complexity of the algorithm, we need the following lemmas.

\begin{lemma} \label{lem:lambda_s_wt_e}
Let $\HGRS$ as defined in Def.~\ref{def:GRS_codes}, $\s = (\c + \e) \HGRS^\top$, $\tau = \wtH(\e)$ and $\alphavec, \ve{v}$ as defined in Corollary~\ref{cor:GVandermonde}. 
\begin{align*}
	\lambda(\s) \leq \tau (\lambda(\e) + \lambda(\HGRS) + 1).
\end{align*}
\end{lemma}

\begin{IEEEproof}
	Let $\Eset = \{i_1,\dots,i_\tau\}$ be error positions and $\e_\Eset=[e_{i_1},\dots,e_{i_\tau}]$ as well as $\H_\Eset = [H_{i_1},\dots,H_{i_\tau}]$, where $H_{i_j}$ is the $i_j$-th column of $\HGRS$. Then, we have $\lambda(\H_\Eset) \leq \lambda(\HGRS)$, $\r \HGRS = \e \HGRS = \e_\Eset \H_\Eset$, and
	\begin{align*}
		\lambda(\s) =& \lambda(\e \HGRS^\top) = \lambda(\e_\Eset \H_\Eset^\top) 
		\leq \tau (\lambda(\e) + \lambda(\H_\Eset) + 1). \: \: \IEEEQEDhere
	\end{align*}
\end{IEEEproof}

Using the better coefficient growth bounds for computations over $\ZZ$ (cf.~Theorem~\ref{thm:recap}), we obtain the following smaller bound in this case.

\begin{lemma} \label{lem:HinZZ}
	Let $\HGRS$ from Def.~\ref{def:GRS_codes} and $\HGRS \in \ZZ^{(n-k)\times n}$ (i.e., $\alpha_i,v_i \in \ZZ$), then
	\begin{align*}
	\lambda(\s) &\leq (\tau+1) \lambda(\e) + \lambda(\HGRS) + n.
	\end{align*}
\end{lemma}

\begin{IEEEproof}
	\begin{align*}
	\lambda(\s) = \lambda(\e \H^\top) =\max\{\lambda(\xi), \lambda((\xi \e) \H^\top)\}+1,
	\end{align*}
	where $\xi = \lcm(\den(e_{i_1}),\dots,\den(e_{i_\tau}))$. Thus, $\xi \e \in \ZZ^n$ and 
	\begin{align*}
	\lambda((\xi \e) \H^\top) \leq \lambda(\xi\e) + \lambda(\H) + (n-1).
	\end{align*}
	Using 
	$\lambda(\xi) \leq \tau \lambda(\e),$
	we obtain
	\begin{align*}
	\lambda(\s) &\leq \max\{(\tau+1) \lambda(\e) + \lambda(\H) + (n-1), \tau \lambda(\e)\}+1 \\
	&= (\tau+1) \lambda(\e) + \lambda(\H) + n. \hspace{3.6cm} \IEEEQEDhere
	\end{align*}
\end{IEEEproof}

\begin{lemma}\label{lem:EEA_cost}
Let $a,b \in \ZZ[x]$ with $d_a = \deg a \geq d_b = \deg b \geq 1$ and $\lambda(a),\lambda(b) \leq t$.
Then, $\mathrm{EEA}(a,b)$ can be computed in
\begin{equation*}
O\!\left( d_a^3 d_b t^2 \log^2(d_a)(\log^2(d_a) + \log^2(t)) \right)
\end{equation*}
bit operations using \cite[Algorithm~6.57, page~188]{von2013modern}.
Furthermore, $\mathrm{EEA}(a,b,t_\mathrm{stop})$ can be computed in 
\begin{equation*}
O\!\left( d_a^3 t^2 \log^2(d_a)(\log^2(d_a) + \log^2(t)) \right)
\end{equation*}
bit operations.
\end{lemma}

\begin{IEEEproof}
The statement follows directly from \cite[Theorem~6.58]{von2013modern}, which is formulated in terms of the $\infty$-norm of the polynomials $a$ and $b$, i.e., $||a||_\infty,||b||_\infty \leq A$ for some integer $A$. This condition is equivalent to $\lambda(a),\lambda(b) \leq \log_2(A)$.
The complexity of the EEA with stopping condition follows from the fact that we save a factor $d_b$ if only one specific triple $(r_h,s_h,t_h)$ in the output sequence of the EEA is explicitly computed, cf.~\cite[page~189]{von2013modern}.
\end{IEEEproof}

In the following, denote by $\den(\alpha)$ and $\num(\alpha)$ the (reduced) denominator and numerator of $\alpha \in \QQ$, respectively.

\begin{lemma}\label{lem:bit width_EEA_input}
Let $S(x)$ be the syndrome polynomial and $\xi = \lcm(\den(s_0),\dots,\den(s_{d-2}))$. 
\begin{align*}
\xi S(x) \in \ZZ[x] \quad \text{ and } \quad 
\lambda(\xi S(x)) \leq d(\lambda(\s)+1).
\end{align*}
\end{lemma}

\begin{IEEEproof}
The first claim, $\xi S(x) \in \ZZ[x]$ is obviously fulfilled (in fact, $\xi$ is the smallest positive integer with this property).
For the bit width, we have
\begin{align*}
\lambda(\xi) &\leq \lambda\left( \prod_{i=0}^{d-2} \den(s_i) \right) \leq \sum_{i=0}^{d-2} \lambda(s_i) + d-1 \\
&\leq (d-1) (\lambda(\s)+1),
\end{align*}
which implies the claim.
\end{IEEEproof}

\begin{lemma}\label{lem:lambda_Lambda}
For  $|\mathcal{E}|=\tau$ and $\Lambda(x)= \prod_{i \in \mathcal{E}} (1-\alpha_i x)= \sum_{j=0}^{t}\Lambda_j x^j$ it holds that
\begin{align*}
\lambda(\Lambda(x)) \leq \tau(\lambda(\alphavec)+2).
\end{align*}
\end{lemma}

\begin{IEEEproof}
Let $\zeta := \prod_{i \in \Eset} \den(\alpha_i)$. Then,
\begin{align*}
\Lambda(x) = \tfrac{1}{\zeta} \prod_{i \in \Eset} \underset{=: \, a_i(x)}{\underbrace{(\den(\alpha_i)-\num(\alpha_i)x)}},
\end{align*}
where $a_i(x) \in \ZZ[x]$, $\deg a_i(x) = 1$, and $\lambda(a_i(x)) \leq \lambda(\alpha_i)$.
By applying Property~\ref{itm:lambda_poly_mult} of Theorem~\ref{thm:recap} inductively, we get
\begin{align*}
\lambda\left(\prod_{i \in \Eset}a_i(x) \right) \leq \sum_{i \in \Eset}\lambda(\alpha_i) +\tau \lambda(2) \leq \tau(\lambda(\alphavec)+2).
\end{align*}
Due to $\zeta \in \NN$ and $\lambda(\zeta) \leq \sum_{i\in\Eset} \lambda(\alpha_i) \leq \tau \lambda(\alphavec)$, the claim follows by $\lambda(\Lambda(x)) \leq \max\{\lambda(\prod_{i \in \Eset}a_i(x)),\lambda(\zeta)\}$.
\end{IEEEproof}

\begin{lemma}\label{lem:lambda_Omega}
For $\Omega(x)= \sum_{j\in \mathcal{E}}e_i v_i \prod_{j \in \mathcal{E}\setminus \{i\}} (1-\alpha_j x)$ and $|\mathcal{E}|=\tau$ it holds that
\begin{align*}
\lambda(\Omega(x)) \leq \tau\big(\lambda(\alphavec)+\lambda(\e)+\lambda(\ve{v})+5\big).
\end{align*}
\end{lemma}

\begin{IEEEproof}
Let $\zeta := \prod_{i \in \Eset} \den(\alpha_i)$, $\eta := \prod_{i \in \Eset}\big( \den(e_i v_i)\big)$ and $a_i(x)\coloneqq\den(\alpha_i) - \num(\alpha_i)x$ then 
\begin{align*}
\Omega(x)= \frac{1}{\zeta\eta}\sum_{i \in \mathcal{E}}\num(e_i v_i) \den(\alpha_i) \prod_{j \in \mathcal{E}\setminus \{i\}} \den(e_j v_j)a_j(x)
\end{align*}
Since $\zeta\eta\cdot\Omega(x)\in \ZZ[x]$ it holds that
\begin{align*}
\lambda(\zeta\eta\cdot\Omega(x))
\leq&\max_{i \in \Eset}\{\lambda\Big(\num(e_i v_i) \den(\alpha_i) \prod_{j \in \mathcal{E}\setminus \{i\}} \den(e_j v_j)\\&a_j(x)\Big)\}+\tau-1
\\
\leq&\max_{i \in \Eset}\{\lambda(\num(e_i))+1+\lambda(\num(v_i))+1 \\&+ \lambda( \den(\alpha_i))+1
+\sum_{j \in \mathcal{E}\setminus \{i\}} \lambda(\den(e_j)\\
&+\tau -1+1+ \sum_{j \in \mathcal{E}\setminus \{i\}} \lambda(\den(v_j)+\tau -1\\
 &+1+\underbrace{\sum_{j \in \mathcal{E}\setminus \{i\}} \lambda(a_j(x))+(\tau-1) \lambda(2)}_{(\tau-1)(\lambda(\alphavec)+2)}\}\\&+\tau-1\\
 \leq& \lambda(\alphavec)+\lambda(\e)+\lambda(\ve{v})+(\tau-1)\lambda(\e)
 \\&+(\tau-1)\lambda(\ve{v})
 +(\tau-1)\lambda(\alphavec)+5\tau
 \\
 =&\tau(\lambda(\alphavec)+\lambda(\e)+ \lambda(\ve{v})+5).
\end{align*}
Also for the bit width of $\zeta \eta$ we have that
\begin{align*}
\lambda(\zeta \eta)= \lambda(\prod_{i \in \Eset}\den(e_i v_i \alpha_i)) \leq  \tau(\lambda(\alphavec)+\lambda(\e)+\lambda(\ve{v})).
\end{align*}
And therefore the claim holds.
\end{IEEEproof}

\begin{lemma}\label{lem:evaluation_complexity}
Let $a(x) \in \QQ[x]$ and $\alpha_i \in \QQ$. Then, the evaluation of $a(x)$ at $\alpha_i$ can be implemented in 
\begin{align*}
	\bigO^\sim(\deg(a(x))\max\{ \lambda(\alpha_i),\lambda(a(x)) \}).
\end{align*}
bit operations\footnote{$\bigO^\sim$ neglects logarithmic terms of its arguments}.
\end{lemma}
\begin{IEEEproof}
Applying Horner's rule~\cite[p.~101]{von2013modern} for evaluation, one gets $\deg a(x)$ multiplications and $\deg a(x)$ additions. By naive computation of the additions and the multiplications as in~\cite{schonhage1971schnelle}, the complexity for integers of bit width $t_1$ and $t_2$ (define: $n_0=\max\{t_1,t_2\}$) is $\bigO(n_0)$, respectively $\bigO(n_0\log n_0 \log\log n_0 )$ (cf.~Algorithms 2.1 in \cite{von2013modern}). Hence,
\begin{align*}
	\bigO(&\deg a(x) n (\log n \log\log  n +1) ) \\
	&=\bigO^\sim(\deg a(x) \max\{ \lambda(\alpha_i),\lambda(a(x)) \} ),
\end{align*}
where $n=\max\{ \lambda(\alpha_i),\lambda(a(x)) \}$.
\end{IEEEproof}

\begin{corollary}\label{cor:forney_complexity}
	For the computation of the $i^\mathrm{th}$ error value $e_i$, $\Lambda'$ and $\Omega$ are evaluated at $\alpha_i^{-1}$ , thus $e_i$, can be computed in
	\begin{equation*}
	\bigO^\sim(\tau^2 (\lambda(\alphavec) + \lambda(\e) + \lambda(\ve{v}))) 
	\end{equation*}
	bit operations.
\end{corollary}

\subsection{Summary of the Algorithm}

\printalgoIEEE{
\DontPrintSemicolon
\KwIn{Received Word $\r = \c + \e$, where $\c \in \CGRS$ and $\wtH(\e) \leq \lfloor \tfrac{n-k}{2} \rfloor$.}
\KwOut{Codeword $\c$}
$\s \gets \r \HGRS^\top$ \;
$S(x) \gets \sum_{i=0}^{d-2} s_i x^i$ \;
$\xi \gets \lcm(\den(s_0),\dots,\den(s_{d-2}))$ \;
$(r_h,s_h,t_h) \gets \mathrm{EEA}(\xi \cdot x^{d-1}, \, \xi \cdot S(x) , \, \tfrac{d-1}{2})$ \label{line:EEA} \;
$c \gets 0^\mathrm{th}$ coefficient of $t_h$ \;
$(\Lambda(x),\Omega(x)) \gets c^{-1} \cdot (t_h,r_h / \xi)$ \tcp*{see below\footnotemark}
$\Lambda'(x) \gets \sum_{i>0} i \Lambda_i x^{i-1}$ \;
$e_i \gets -\tfrac{\alpha_i}{v_i} \tfrac{\Omega(\alpha_i^{-1})}{\Lambda'(\alpha_i^{-1})}$ for $i=1,\dots,n$ \;
\Return{$\c = \r-\e$}
\caption{Decoding Algorithm for GRS Codes over $\QQ$}
\label{alg:DecodeGRS}
}
\footnotetext{Note that this was stated wrongly in the short version~\cite{sippel2019RSchar0}.}

\begin{theorem}\label{thm:decoding_main_statement}
Algorithm~\ref{alg:DecodeGRS} returns the correct codeword if $\wtH(\e)\leq \tfrac{n-k}{2}$. Its complexity in bit operations is
\begin{equation*}
\bigO^\sim\!\Big( d^7 \big[\lambda(\e)+\lambda(\HGRS)\big]^2 +n^4[\lambda(\c) + \lambda(\e) +\lambda(\HGRS) ]\Big). 
\end{equation*}
\end{theorem}

\begin{IEEEproof}
Correctness follows directly by Lemmas~\ref{lem:EEA_key_eq} and \ref{lem:forney} above and the computational complexity of $\s$ in Line~1. Since the latter is a usual vector-matrix-multiplication and the computation itself depends on the codeword $\c$, its complexity is bounded by 
\begin{align*}
\bigO\!\left(n^4[ \lambda(\c) + \lambda(\e) + \lambda(\HGRS) ]\right).
\end{align*}
The remaining part of the algorithm is independent of the codeword $\c$.
The second bottleneck of the algorithm is the EEA, which can be implemented by \cite[Algorithm~6.57, page~188]{von2013modern}.
The input polynomials, $a(x) = \xi \cdot x^{d-1}$ and $b(x) = \xi \cdot S(x)$ have bit width
\begin{align*}
\lambda(a(x)),\lambda(b(x)) &\leq d(\lambda(\s)+1) \leq d\tau(\lambda(\e)\!+\!\lambda(\HGRS)\!+\!2) \\
&\in \bigO\Big(d^2\big[\lambda(\e)+\lambda(\HGRS)\big]\Big)
\end{align*}
by Lemmas~\ref{lem:bit width_EEA_input} and~\ref{lem:lambda_s_wt_e}, and degrees $d_a = d-1 $ and $d_b \leq d-2$.
By Lemma~\ref{lem:EEA_cost}, we can therefore bound the cost of Line~\ref{line:EEA} by
\begin{align*}
\bigO\!\Big( d^7 &\big[\lambda(\e)+\lambda(\HGRS)\big]^2 \cdot \\
 &\log^2(d)\log^2\Big(d^2\big[\lambda(\e)+\lambda(\HGRS)\big]\Big) \Big).
\end{align*}
The remaining steps of Algorithm~\ref{alg:DecodeGRS} are negligible compared to this step due to Corollary~\ref{cor:forney_complexity}.
\end{IEEEproof}

\begin{remark}
	The syndrome computation complexity depends on $\lambda(\c)$, which itself depends on the bit width of the information $\info$. We can bound $\lambda(\c)$ depending on $\lambda(\info)$ by Theorem~\ref{thm:lambda_c}. 
\end{remark}
For a special choice of $\alphavec$, the following corollary holds. We state it in terms of $n$ as we often have $d \in \Theta(n)$.

\begin{corollary}\label{cor:final_complexity}
If the error $\e$ has bit width at most $t$, codeword $\c$ at most $t'$ and the code locators are chosen with $\lambda(\alphavec) \in \bigO(\log(n))$ (e.g., $\alpha_i = i$ for $i=1,\dots,n$), then Algorithm~\ref{alg:DecodeGRS} can be implemented in
\vspace{-0.05cm}
\begin{equation*}
\bigO^\sim\!\left( \max\{n^7t^2,n^9,n^4 t'\} \right)
\vspace{-0.05cm}
\end{equation*}
bit operations.
In case of $\HGRS \in \ZZ^{(n-k)\times n}$ we get
\vspace{-0.05cm}
\begin{equation*}
	\bigO^\sim\!\left( \max\{n^7t^2,n^7,n^4 t'\} \right) = \bigO^\sim(\max\{n^7t^2, n^4 t'\}).
\vspace{-0.05cm}
\end{equation*}
\end{corollary}

\section{Trade-Off between Encoding and Decoding}
\begin{table*}
	\caption{Upper bounds for the bit widths of codeword $\c$ and syndrome $\s$ for several choices for $v_i$ and $v_i'$.  
	} \label{tab:choicesforsyndromandcodeword}
	\vspace{-0.5cm}
	\begin{center}
		\begin{tabular}{p{2cm}p{4.1cm}p{4.6cm}p{5cm}}
			\toprule
			general 			& \multicolumn{2}{c}{$ \lambda(\c) \leq k(\lambda(\info) + \lambda(\G) + 1)$} 
			& $\lambda(\ve{s}) \leq \tau(\lambda(\e) + \lambda(\HGRS) +1)$ \\
			\midrule
			& $\G = \GGRS$ & $\G=\G_\mathrm{Cauchy}$ & \\
			\cmidrule{2-3}
			$c_id_j=1$			& $k(\lambda(\info) + 3(k-1)\lambda(\alphavec) + k)$ &$k(\lambda(\info) + 2\lambda(\alphavec) + 2 )$
			& $\tau(\lambda(\e) + (3n-3k-1)\lambda(\alphavec) + n-k+1)$ \\
			
			$v_i=1$, $v_i'=w_i$	& $k(\lambda(\info) + (2n+k-3)\lambda(\alphavec)+n)$  &$k(\lambda(\info) + (2n-2k+1)(2\lambda(\alphavec) + 1) +1)$ 
			& $\tau (\lambda(\e)  + (n-k-1)\lambda(\alphavec) +  1) $ \\
			\bottomrule
		\end{tabular}
	\end{center}
	
\end{table*}
Corollary~\ref{cor:final_complexity} states that the complexity might be reduced by choosing the entries of $\HGRS$ in $\ZZ$. The next corollary also compares this reduction.

\begin{corollary}
	Regard the setting of Lemma~\ref{lem:lambda_s_wt_e}.
	\begin{itemize}
		
		\item 	For the case of $v_i = 1 \, \forall i=1,\dots,n$, we have
		\begin{align*}
		\lambda(\s) \leq \tau (\lambda(\e)  + (n-k-1)\lambda(\alphavec) +  1).
		\end{align*}
		\item For the case $c_id_j = 1$, we have
		\begin{align*}
		\lambda(\s) \leq \tau(\lambda(\e) + 3(n-k)\lambda(\alphavec) + n-k+1).
		\end{align*}
		\item 	
		For the case of $\HGRS \in \ZZ^{(n-k)\times n}$ and $v_i = 1 \, \forall i=1,\dots,n$, we have
		\begin{align*}
		\lambda(\s) \leq (\tau +1)\lambda(\e)  + (n-k-1)\lambda(\alphavec) +  n.
		\end{align*}
		\item For the case $\HGRS \in \ZZ^{(n-k)\times n}$ and $c_id_j = 1$, we have
		\begin{align*}
		\lambda(\s) \leq (\tau+1)\lambda(\e) + 3(n-k)\lambda(\alphavec) + 2n-k.
		\end{align*}
	\end{itemize}
\end{corollary}
\begin{IEEEproof}
	Insert the results from Table~\ref{tab:choicesv} into $\lambda(\HGRS)$.
\end{IEEEproof}

In this case, however, we must have $v_i \in \ZZ$, which might yield large values of $\lambda(\G)$ according to Theorem~\ref{thm:closed_form_v_iv_i'}. 
Note that there might be alternative generator and parity check matrix combinations with smaller overall coefficient growth. 
This again can increase the coefficient growth during encoding. Hence, one can find a tradeoff between decoding complexity and coefficient growth during encoding by choosing $v_i$.

Table~\ref{tab:choicesforsyndromandcodeword} shows a part of the tradeoff by comparing the upper bounds for the bit width of syndrome and codeword, that are derived in Theorem~\ref{thm:lambda_c} and Lemma~\ref{lem:lambda_s_wt_e} in the sections before. 
Since $\lambda(\alphavec) \in \bigO(\log n)$, we get $\lambda(\c) \in \bigO(n^2\log n)$ for all cases except for the Cauchy generator matrix $\G_\mathrm{Cauchy}$ in combination with $c_id_j=1$, where we have $\lambda(\c)\in \bigO(n\log n)$. Hence, this choice is asymptotically the best, since the syndrome has $\lambda(\s) \in \bigO(n^2\log n)$ anyway.

\section{Conclusion}

In this paper we studied RS codes over arbitrary fields and analyzed the coefficient growth for encoding and decoding over the rational numbers.
By deriving bounds on the intermediate coefficient growth during computations, we were able to show that decoding up to half-the-minimum distance is possible in a polynomial number of bit operations in the code length and bit width of the error values.

As also mentioned by the reviewers, we believe, that this paper is a starting point for research on RS codes over number fields and there are many open problems that need to be studied in future work. 
For instance, we can consider other decoding algorithms such as Berlekamp-Welch, Berlekamp-Massey or list decoding approaches. 
Further, the results can be extended to a wider class of number fields. 
Also reduction of the computation modulo a prime by decomposing the number field into prime ideals as in~\cite{augot2017generalized} is a possibility for further research. 
Another important aspect is to study possible applications in detail. 
Moreover, the methods can be used to determine the bit complexity of the algorithms in \cite{robert2016quadratic,muelich2016alternative} for decoding Gabidulin codes over characteristic zero.

\section{Appendix}
Proofs for formulas in Table~\ref{tab:choicesv}.
Case: $c_id_j=1$, then \\
$\lambda(\GGRS) \leq (k-1)(3\lambda(\alphavec) + 1)$
\begin{IEEEproof}
	\begin{align*}
	\lambda(\ve{v}') \leq& \max\limits_{i=1,\dots,n} \{ \lambda(\prod_{\substack{1\leq t\leq k\\t \neq i}} (\alpha_i - \alpha_t) )\} \\
	\leq& \max\limits_{i=1,\dots,n} \{ \sum_{\substack{t=1,\dots,k\\t\neq i}}(\lambda(\alpha_i) + \lambda(\alpha_t) +1) \} \\
	\leq& (k-1)\max\limits_{i=1,\dots,n} \{ 2\lambda(\alpha_i) + 1 \} \\
	=& (k-1)(2\lambda(\alphavec) + 1)
	\end{align*}
\end{IEEEproof}

Case: $v_i = 1, v_i'=w_i$, then \\
$\lambda(\GGRS) \leq (2n+k-3)\lambda(\alphavec) + n-1$
\begin{IEEEproof}
	\begin{align*}
		\lambda(\ve{v}') =& \max\limits_{i=1,\dots,n} \{ \lambda( \prod_{\substack{1\leq t\leq n\\t \neq i}} (\alpha_i - \alpha_t)^{-1} ) \} \\
		\leq& \max\limits_{i=1,\dots,n} \{ \sum_{\substack{t=1,\dots,n\\t\neq i}} (\lambda(\alpha_i) + \lambda(\alpha_t) + 1) \} \\
		\leq& (n-1)(2\lambda(\alphavec) + 1)
	\end{align*}
\end{IEEEproof}

Case: $v_i=1, v_i'=w_i$, then \\
$\lambda(\G_\mathrm{Cauchy}) \leq (2n-2k+1)(2\lambda(\alphavec) + 1)$
\begin{IEEEproof}
	Inserting the condition into $\A$ gives 
	\begin{align*}
		A_{ij} =& (\alpha_i - \alpha_{j+k})^{-1} \prod_{k+1\leq t\leq n} (\alpha_{j+k}-\alpha_t)^{-1} \\ &\cdot \prod_{k+1\leq t\leq k} (\alpha_i - \alpha_t) .
		\end{align*}
	\begin{align*}
		\lambda(\G_\mathrm{Cauchy}) \leq& \max\limits_{\substack{i=1,\dots,k\\j=1,\dots,n-k}} \{ \sum_{t=k+1,\dots,n} \lambda(\alpha_{j+k} - \alpha_t) \\
		&+ \sum_{t=k+1,\dots,k} \lambda(\alpha_i - \alpha_t) + \lambda(\alpha_i) + \alpha_j + 1 \} \\
		\leq& 2(n-k)(2\lambda(\alphavec) + 1) + 2\lambda(\alphavec) + 1
	\end{align*}
\end{IEEEproof}

Case: $v_i'=1, v_i=w_i$, then \\
$\lambda(\G_\mathrm{Cauchy}) \leq 2(k-1)(2\lambda(\alphavec) + 1)$
\begin{IEEEproof}
	\begin{align*}
		\lambda(\G_\mathrm{Cauchy}) \leq& \max\limits_{\substack{i=1,\dots,k\\j=1,\dots,n-k}} \{ \lambda(\prod_{\substack{1\leq t\leq k\\t \neq i}} (\alpha_{j+k}-\alpha_t) \\
		&\cdot \prod_{\substack{1\leq t\leq k\\t \neq i}} (\alpha_i - \alpha_t)^{-1} ) \} \\
		\leq& (k-1)(2\lambda(\alphavec)+1) + (k-1)(2\lambda(\alphavec)+1)
	\end{align*}
\end{IEEEproof}

Case: $v_i'=1$, $v_i=w_i$, then \\
$\lambda(\HGRS) \leq (3n-k-3)\lambda(\alphavec) + n-1$
\begin{IEEEproof}
	\begin{align*}
		\lambda(\ve{v}) \leq& \max\limits_{i=1,\dots,n} \{ \lambda(\prod\limits_{\substack{t=1,\dots,n\\t\neq i}} (\alpha_i - \alpha_t) ) \}\\
		\leq& (n-1)(2\lambda(\alphavec) + 1)
	\end{align*}
\end{IEEEproof}

Case $c_id_j = 1$, then \\
$\lambda(\HGRS) \leq (3n-3k-1)\lambda(\alphavec) + n-k$
\begin{IEEEproof}
	Representation of $v_i$, see Remark~\ref{rem:c_id_j=1}.
	\begin{align*}
		\lambda(\ve{v}) \leq& \max\limits_{i=1,\dots,n} \{ \lambda(\prod_{1\leq t\leq n-k} (\alpha_{i} - \alpha_{t+k}) ) \} \\
		\leq& (n-k)(2\lambda(\alphavec)+1)
	\end{align*}
\end{IEEEproof}

\bibliographystyle{IEEEtran}
\bibliography{main}

\end{document}